\documentclass[%
 aip,
 amsmath,amssymb,
 reprint,%
]{revtex4-1}

\usepackage{graphicx}
\usepackage{dcolumn}
\usepackage{bm}
\clearpage
\usepackage[utf8]{inputenc}
\usepackage[T1]{fontenc}
\usepackage{mathptmx}
\usepackage{float}
\usepackage{xcolor}
\newcommand{\bc}{B$_4$C }
\newcommand{\bcend}{B$_4$C}

\begin{document}

\title{Physical and Thermoelectric Properties of 2D B$_4$C Nanosheets} 

\author{Adway Gupta, Tathagata Biswas, Arunima K. Singh}

\affiliation{Department of Physics, Arizona State University, Tempe, Arizona  85287-1504}

\date{\today}

\begin{abstract}
 Boron carbide (\bcend) has been well studied both theoretically and experimentally in its bulk form due to its exceptional hardness and use as a high temperature thermoelectric. However, the properties of its two-dimensional nanosheets are not well established. In this paper, using van der Waals corrected density-functional theory (DFT) simulations, we show that the bulk \bc can be cleaved along different directions to form \bc nanosheets with low formation energies. We find that there is minimal dependence of the formation energies on the cleavage planes and surface terminations. Whilst the density of states of the bulk \bc indicate that it is a semiconductor, the \bc nanosheets are found to be predominantly metallic. We attribute this metallic behaviour to a redistribution of charges on the surface B-C bonds of the films. The Seebeck coefficients of the the \bc films remain comparable to those of the bulk, and are nearly constant as a function of temperature. Our results provide guidance for experimental synthesis efforts and future application of B$_4$C nanosheets in nanoelectronic and thermoelectric applications. \\ 
\end{abstract}


\maketitle 

 Boron rich compounds have garnered significant scientific attention over the last several decades.\cite{boronrich1,boronrich3,boronrich4} Boron, unlike its neighbour on the periodic table, carbon, has decidedly unusual covalent bonding patterns. Even in the simplest multi boron compound, diborane(B$_2$H$_6$), the "banana" bond between the boron orbitals hybridised with those of hydrogen, leave only 2 electrons per bond to be shared between 3 atoms.\cite{diborane1} These unusual bonding patterns often result in functional physical and thermoelectric properties in boron rich compounds, for instance, in boron carbide(\bcend).\cite{unusualboron,unusualboron2}\\ 
Bulk \bc is an excellent high temperature thermoelectric with Seebeck coefficients between 200 to 400 $\mathrm{\mu}$V/K, that remain almost invariant from 300 K up to 1500 K. It has large electrical conductivity ($10^3$ to $10^4$ $\Omega^{-1} \mathrm{m}^{-1}$) at reasonable carrier concentrations.\cite{thermoelectric1,thermoelectric2} Finally, it is also one of the hardest known materials with a hardness of 9.3 on the Mohs scale,\cite{b4chard} making it harder than steel (8 Mohs), and softer than diamond (10 Mohs).\cite{hardnesshandbook} It is also semiconducting, with an experimentally measured bandgap of 1.6 \textendash \hspace{0.05cm} 2 eV.\cite{bandgap1,bandgap2,bandgap3,bandgap4}  
 
These interesting properties have led to extensive studies of \bcend. Figure 1 (a) and (b) show the rhombohedral unit cell of \bc with a C-C-C chain along the body diagonal and with the 20-atom boron icosahedra at the vertices of the unit cell. Due to the lack of electrons needed to form pairwise boron-boron bonds within the icosahedra, the boron atoms are bonded in a unique bonding scheme, called triple center bonding.\cite{b4cbonding} All the atoms in \bc are strongly bound and there are no weak van der Waals (vdW) bonds which could be easily broken to form low dimensional structures like nanosheets via exfoliation techniques. Thus, most of the work in generating two-dimensional (2D) boron carbide nanostructures has employed bottom up approaches like CVD on a variety of substrates.\cite{cvd,b4cthinfilms1} 
\begin{figure}[h]
    \centering
    \includegraphics[width=3in]{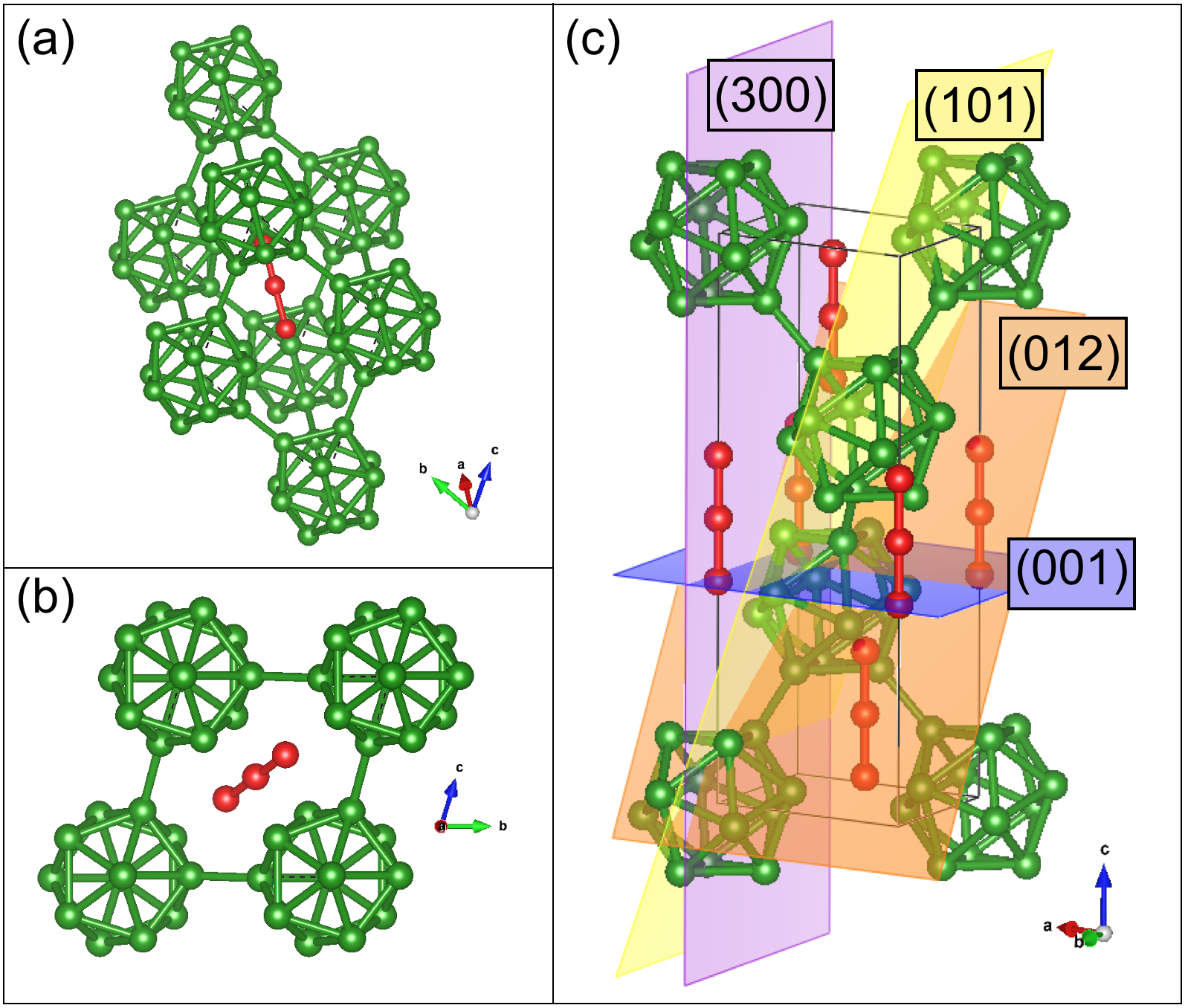}
    \caption{(a) and (b) show the \bc primitive cell projected along two different directions. The red and green spheres denote carbon and boron atoms, respectively. In (c), four cleavage planes of \bc are shown in the conventional cell of \bcend. The (300) and (001) planes cleave the \bc parallel and perpendicular to the C-C-C chains whereas (101) and (012) cleave along directions that are non-orthogonal to both the cell and the C-C-C chains.}
    \label{B4CStruct}
\end{figure}

In a recent article,\cite{prevpaper} using both experiments and first-principles simulations, we have shown that 2D \bc nanosheets can be unexpectedly synthesized using a variety of top-down synthesis techniques. Figure 1 (c) shows the orientations of four planes along which \bc can be cleaved to create nanosheets. We considered all the possible surface atomic terminations for each cleavage plane for a total of 28 unique terminations. We then computed the formation energy for each plane and termination using density functional theory(DFT).
In the existing literature, formation energies have been routinely used as a heuristic measure of synthesizability of 2D materials. \cite{nanosheetthreshold} 2D materials with formation energies lower than 0.2 eV/atom are likely to be stable and extractable by exfoliation techniques. We found that the formation energies along all directions are similar, indicating no preferential direction of cleavage. Additionally, the four minimum formation energies were found to be 0.113, 0.056, 0.116, and 0.105 eV/atom along the (300), (100), (101), and (012) cleavage planes, respectively, which are all lesser than the aforementioned threshold of 0.2 eV/atom. We confirmed these DFT predictions through experiments, where after a mild sonication of bulk \bc in a variety of organic solvents we obtained 2D flakes oriented along a range of arbitrary directions.\cite{prevpaper}
Our \bc nanosheet formation energy results\cite{prevpaper} were surprising since \bc has no weak vdW bonds that are typically broken to form low-dimensional forms of materials. In order to gain insight and determine the origin of this seemingly anomalous behavior, we examined the structures of the near-surface atoms of the nanosheets, and found a rearrangement of the surface atoms into smaller cage structures.

In this article, using DFT and excited state theory $G_0W_0$ simulations, we provide insight into the low formation energies of the nanosheets of this non-vdW bonded material. We examine the bonding and charge distribution of all the nanosheets to explain the structure of the near-surface atoms. We also compute their electronic and thermoelectric properties to guide the future application of \bc nanosheets in nanoelectronics, thermoelectric devices, and as reinforcement materials.

All the density functional theory (DFT) simulations are based on the projector-augmented wave method as implemented in the plane-wave code VASP.~\cite{Kresse5, Kresse4, Kresse3} DFT simulations were performed using the vdW-DF-optB88 exchange–correlation functional,~\cite{Klimes2011} with a cutoff energy for the plane wave basis at 600 eV, a $k$-point density of 60 \AA$^{-1}$ in the $x$ and $y$ directions and only 1 $k$-point in the $z$ direction for all the slab simulations. Structural relaxations have been performed with energy convergence within $10^{-6}$ eV in each ionic step and till the forces are converged to 0.005 eV/atom. The quasiparticle (QP) energies were obtained by using many body perturbations theory within the G$_0$W$_0$ approximation for the self-energy operator. \cite{shishkin2006implementation,shishkin2007self} The basis set size for the response functions and screened Coulomb potential, $W$, was chosen to include all plane waves up to an energy cutoff of 250 eV. The number of unoccupied bands included in the $GW$ calculation was set to 160 such that the QP gap is converged to within 0.1 eV. For the frequency integral in the calculation of self-energy we have used 80 frequency points. To obtain the QP bandstructure we have used interpolation formalism implemented in WANNIER90 package.\cite{mostofi2008wannier90}

\begin{figure}
    \centering
    \includegraphics[width=3in]{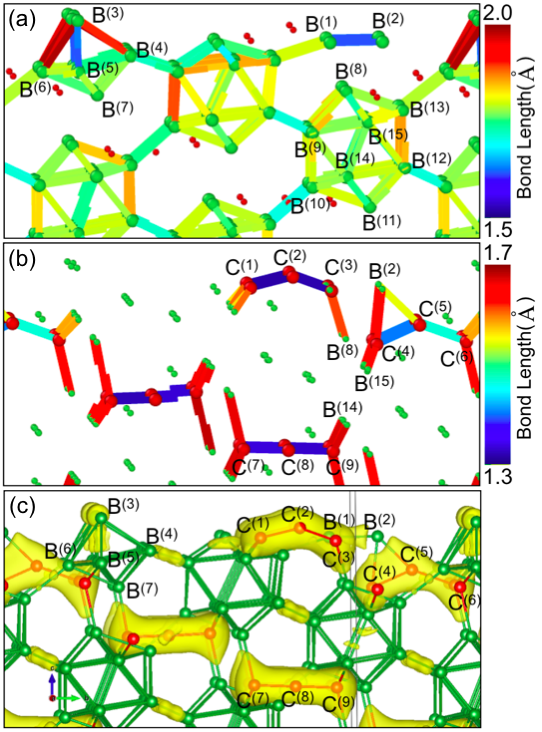}
    \caption{(a) Bond length map for B-B bonds of \bc cleaved along the (300) plane and a particular termination. (b) Bond length map for B-C and C-C bonds for the same nanosheet projected along the same plane. (c) Charge density isosurface with saturation levels set to 0.3 $e$/\AA$^3$ to futher highlight the bonding schemes. }
    \label{Figure3}
\end{figure}

\begin{figure*}[htp]
    \centering
    \includegraphics[width=0.85\linewidth]{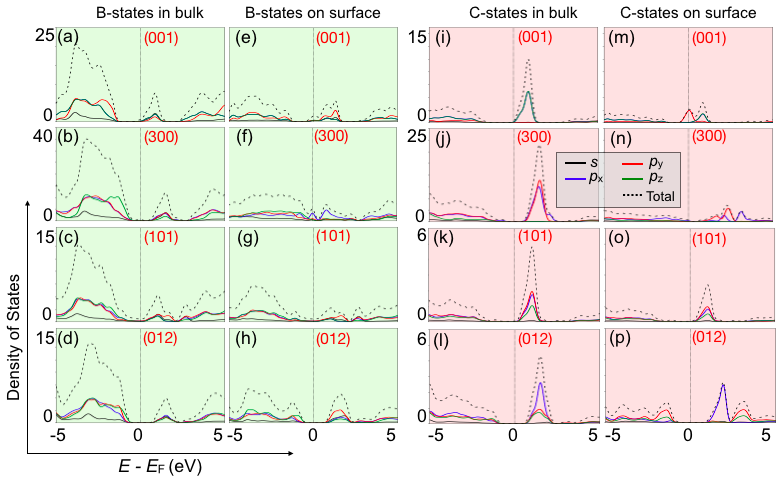}
    \caption{Figure shows the DFT computed total and element-orbital projected density of states for the minimum energy terminations of the \bc nanosheets. (a)-(d) shows state of B atoms in the bulk, (e)-(h) that of B atoms on the surface, (i)-(l) that of C atoms in the bulk, and (m)-(p) that of  C atoms on the surface. Clearly, the nature of the DOS is similar for all the bulk atoms, while for the surface atoms, several states can be seen at the Fermi level.}
    \label{slabgap}
\end{figure*}

\begin{figure}[h]
    \centering
    \includegraphics[width=\linewidth]{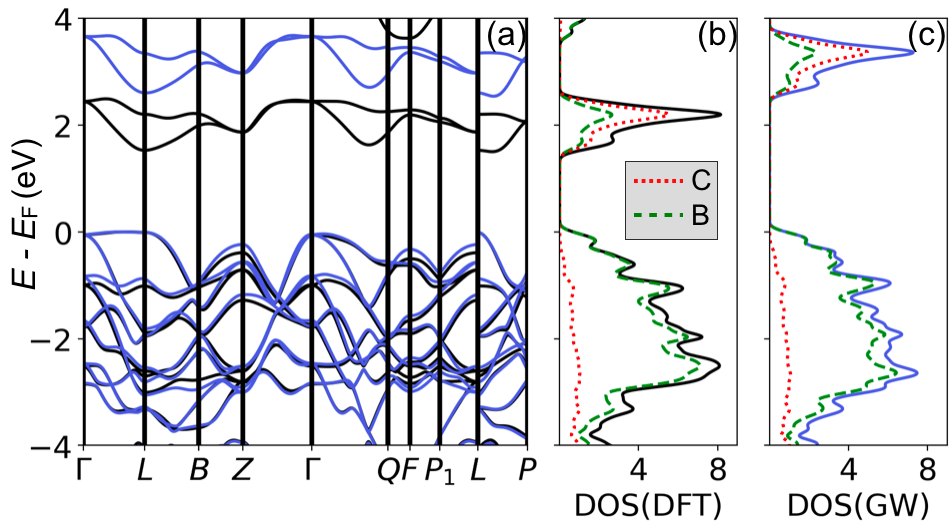}
    \caption{The DFT (black lines) and $G_0W_0$ (blue lines) computed bandstructures of bulk \bc are shown in (a). The DFT computed total and partial density of states of boron and carbon atoms are shown in (b) and (c) shows the $G_0W_0$ computed states.  }
    \label{bulkgap}
\end{figure}

\begin{figure*}
    \centering
    \includegraphics[width=\linewidth]{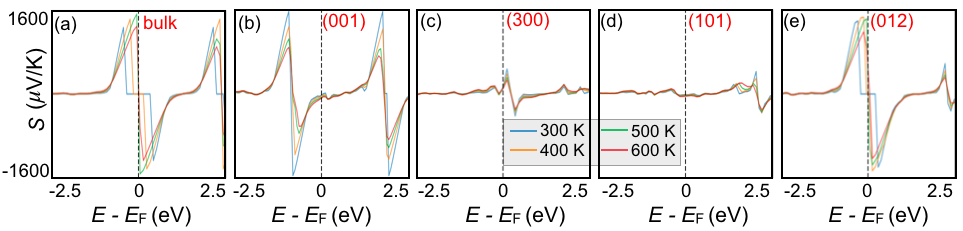}
    \caption{The Seebeck coefficients of (a) bulk \bcend, and the minimum energy terminations of 2D \bc nanosheets cleaved along (b) (001), (c) (300), (d) (101), and (e) (012) plotted in an energy window of $\pm$ 2.5 eV around the Fermi energy.}
    \label{Sfigure}
\end{figure*}

We start by examining the bonding and charge densities of the nanosheets. Figure~\ref{Figure3} (a) and (b) show the B-B bonds, B-C bonds, and the C-C bond of the near-surface atoms of one of the terminations along the (300) plane. The color bars denote the lengths of the bonds. Figure~\ref{Figure3}(c) shows the charge density isosurface of the same structure normalised to a saturation level of 0.3 $e$/\AA$^3$ such that the charge densities on atoms are clearly defined. \cite{vesta} The boron(carbon) atoms are labelled as B$^{(n)}$(C$^{(n)}$) where ($n=1,2,..$), so that individual atoms can be clearly referenced in the discussion. All the 3 plots are constructed along the same projection. We note that most boron atoms at the surface, for example B$^{(3)}$-B$^{(7)}$ in Figure~\ref{Figure3} (a), rearrange into smaller cage structures with atoms fewer than that in the icosahedra of the bulk \bcend. The B-B bond for these boron atoms are partly larger ($\sim$ 2.0 \AA) and partly smaller ($\sim$ 1.5 \AA) compared to those in the icosahedra of bulk-like atoms (for example, $\sim$ 1.7 \AA~bond length of B$^{(8)}$-B$^{(15)}$ that are located far from the surface). However, examination of the charge density at these sites in Figure~\ref{Figure3} (c) reveals that there is a significant charge transfer between the B atoms, suggesting an existence of an ionic bond between B$^{(3)}$ and B$^{(5)}$, and strong covalent bonds between the other B atoms in the cage. We find that some of the surface B atoms, like B$^{(1)}$ and B$^{(2)}$ in Figure~\ref{Figure3} (a), have smaller bond-lengths than the B atoms in the bulk, and thus are strongly bound.

We find that C-C bond lengths at the surface of the nanosheets to be similar to those far from the surface, i.e., similar to the C-C bond lengths in bulk \bcend. For instance in Figure~\ref{Figure3} (b), the C$^{(4)}$-C$^{(5)}$-C$^{(6)}$ bond lengths are comparable to C$^{(7)}$-C$^{(8)}$-C$^{(9)}$ bond lengths. However, the surface C chains are bent in contrast to the straight C chains in bulk \bcend. The bending of the C chains can be explained by inspecting the charge density distribution, for instance between B$^{(2)}$, C$^{(4)}$ and C$^{(5)}$ in Figure~\ref{Figure3} (c). We see there that there is charge accumulation between the B$^{(2)}$ and the C$^{(4)}$ and C$^{(5)}$ atoms, indicating a strong bonding of the C chains with solitary boron atoms at the surface \textemdash this holds the surface boron atoms to the nanosheet structure. In comparison, the bend in the surface C$^{(1)}$-C$^{(2)}$-C$^{(3)}$ compared to the bulk C$^{(7)}$-C$^{(8)}$-C$^{(9)}$ can simply be explained by the asymmetry of B atoms on the top and bottom of the chain. Notably, the overall charge distribution on the C chains is unaltered due to the creation of the surface of the nanosheet which explains the nearly constant C chain bond lengths at and far from the surface of the nanosheets. Additionally, for atoms near the C chains, we see that the electron density is larger on the C atoms than the B atoms, implying that the bonds between the C-chains and the icosahedra are ionic in nature. Hence the inter-icosahedra bonds are just as strong, perhaps stronger, than the intra-icosahedra bonds. This unique bonding scheme leads to an extremely stable structure, both in the bulk \bc and at the atoms far away from the surface of nanosheets. The bond length maps of all other nanosheets (available in the Supplementary Materials) show a similar trend. Thus we establish that the nanosheets are highly stable, with strong bonds at and far from the surface of the nanosheets explaining their low formation energies.

Next, we calculated the electronic properties of \bc nanosheets. Figure~\ref{slabgap} shows the element-projected electronic density of states (DOS) for the \bc nanosheets computed from DFT. The DOS belonging to the B-atoms and the C-atoms are shown by green and red shaded sub-figures, respectively. The figure shows the DOS of the lowest formation energy termination for each cleavage direction, i.e. (001), (300), (101), and (012). The DOS of all the other nanosheets studied in this work can be found in the Supplementary Materials. We find that there are states at the Fermi level indicating that nearly all of the \bc nanosheets are metallic in nature. 

In contrast, the bulk \bc is a known semiconductor. \cite{bandgap1, bandgap3, b4cgapexp} Figure~\ref{bulkgap} compares the $G_0W_0$ and DFT computed bandstructure and the density of states of the bulk \bcend. The DFT computed band gap is found to be 1.53 eV, similar to the theoretically computed bandgaps reported in the literature.\cite{bandgap3} The $G_0W_0$ bandgap is found to be 2.3 eV, comparable to the experimentally measured values of 2.09 eV reported by Werheit.\cite{b4cgapexp} $G_0W_0$ simulations are often an order of magnitude more expensive than standard DFT simulations but they are found to be remarkably successful in predicting bandgaps of semiconducting and insulating materials. It is important to mention here that we are reporting results from the first $G_0W_0$ calculations performed on bulk B$_4$C. 

Coming back to the \bc nanosheets, defining the surface-atoms as the atoms within the top and bottom 15 \% of the total slab thickness, we project the DOS of the bulk-like and surface-like atoms separately in the  Figure~\ref{slabgap}. We can see that the DOS of the bulk-like atoms remain similar to that of the bulk \bc states for all the lowest energy terminations of the four cleavage planes. This result is expected as the bonding of the atoms in the far-from surface of the nanosheets closely resembles the bonding in bulk \bcend. The Fermi level, i.e. the valence band maxima energies, are slightly different for the different planes as the nanosheets deviate nominally from the B-C ratio of 4:1 for the various plane terminations. The states attributed to the surface of the nanosheets have a drastically different nature than that of the bulk-like atoms. Figure~\ref{slabgap} shows that for most terminations there are now states at the Fermi level. These additional states are contributed by both boron and carbon atoms. In the process of rearrangement of atoms close to the cleaved surface, the band structure of the material changes leading to new hybridized states near the Fermi level. Thus 2D \bc shows metallic character on its surface. The emergence of similar surface metallicity as a result of surface reconstruction has been observed in other materials as well.\cite{duke1996semiconductor}

To understand the impact of reduction in the dimensionality of \bc on its thermoelectric properties we compute the Seebeck coefficient of the \bc nanosheets. Thermoelectric materials can directly convert waste heat into electrical energy based on the Seebeck effect. The Seebeck coefficient ($S$) is given by Eq \ref{Seq}.
\begin{equation}\label{Seq}
     S = -\frac{\Delta V}{\Delta T}
\end{equation}
where $\Delta V$ and $\Delta T$ are the voltage and temperature differences across the material, respectively. The energy conversion efficiency of a thermoelectric material is proportional to the square of its Seebeck coefficient. We estimate the Seebeck coefficients of the \bc nanosheets the using the BoltzTrap2 code, \cite{madsen2018boltztrap2} which solves the semiclassical Boltzmann transport equation (BTE) under the constant relaxation time approximation. 

The Seebeck coefficients of bulk \bc and the minimum energy terminations for the 4 chosen directions of cleavage plotted against the energy are shown in Fig \ref{Sfigure}. Notably, the Seebeck coefficient for all the nanosheets remain almost invariant across the temperature range of 300-600 K. This is an encouraging property for its use as a high temperature thermoelectric by increasing the effective operating temperatures of the material. The Seebeck coefficients for the minimum energy terminations along (001) and (012) are similar in magnitude to that of bulk \bcend, which itself is a widely used thermoelectric material at appropriate carrier concentrations, implying that these nanosheets are equally good thermoelectric materials. However, the $S$ values for the terminations along (300) and (101) are about 75\% smaller than that of the bulk \bcend. We examine the semiclassical BTE to understand this deviation. \\

 The semiclassical BTE computes physical parameters of a macroscopic system by expressing them in terms of specific moments $\mathcal{L}^{(\alpha)}$, which emerge from the mathematics of the BTE formalism. \cite{ashcroft} For example, the electrical current, $j$, is given by
\begin{equation}
    j = \Delta V \mathcal{L}^{(0)} + \frac{\Delta T}{eT} \mathcal{L}^{(1)}
\end{equation}
where $e$ is the electronic charge, $T$ the  temperature of the system, and $\mathcal{L}^{(0)/(1)}$ are the aforementioned moments. In the situation where $j=0$, We can represent the Seebeck coefficient using Eq \ref{Seq} as,
\begin{equation}\label{Seq2}
    S = -\frac{\Delta V}{\Delta T} = \frac{\mathcal{L}^{(1)}}{eT\mathcal{L}^{(0)}}
\end{equation}
The $\alpha^{th}$ moment, $\mathcal{L}^{(\alpha)}$, is calculated by summing over all the bands, $n$, of integrals over the entire reciprocal $k$-space, given by Eq \ref{int},
\begin{equation}\label{int}
    \mathcal{L}^{(\alpha)} = e^2\sum _ n \int \mathbf{d^3k}(\epsilon_n(k)-\mu)^\alpha \begingroup\color{gray}\Big(\frac{\partial \epsilon_n(k)}{\partial k}\Big)^2\endgroup \tau_n(k)\frac{\partial g^0}{\partial \epsilon_n}
\end{equation}
where, $\mu$, $e$, and $\tau_n$ are the chemical potential, electronic charge, and the relaxation time in the $n^{th}$ band, respectively. The $g^0$ is the Fermi distribution function and $\epsilon_n$ is the energy relation for the $n^{th}$ band. The term of significance for us, highlighted in gray in  Eq \ref{int}, is the derivative of the energy with respect to the reciprocal point $k$. It clearly changes the value of the moment $\mathcal{L}^{(1)}$ and in turn of $S$ given by Eq \ref{Seq2}. Thus, materials with drastically different features in the energy bandstructure near the Fermi level, will have drastically different Seebeck coefficients. Supplementary Figure 5 shows the bandstructures for the four minimum energy terminations. A comparison between these plots and the bandstructure of the bulk, shown in Fig \ref{bulkgap}, clearly illustrates that the bandstructures for the bulk, (001), and (012) are very similar to each other and different from those of (300) and (101) terminated nanosheets. We can conclude that the differences in the Seebeck coefficients for the nanosheets are a result of the differences in the curvature of bands near the Fermi level in their respective bandstructures. 

In summary, in this work we have established that 2D \bc nanosheets are energetically stable due to a rearrangement of bonds near the cleaved surfaces. Additionally, we show that they exhibit promising properties like a semiconductor-metal transition due to the reduction in dimensionality, while also maintaining the encouraging thermoelectric properties of it's bulk counterpart. 

\begin{acknowledgments}
The authors acknowledge the San Diego Supercomputer Center under the NSF-XSEDE Award No. DMR150006 and the Research Computing at Arizona State University for providing HPC resources. This research
used resources of the National Energy Research Scientific Computing Center, a DOE Office of Science User Facility supported by the Office of Science of the U.S. Department of Energy under Contract No. DE-AC02-05CH11231. AG and AS acknowledge support by NSF DMR under grant \# DMR-1906030. Authors thank Q.H. Wang for fruitful discussions.
\end{acknowledgments}

\section*{Data Availability Statement}
The data that support the findings of this study are available from the corresponding author upon reasonable request.

\section{References}

\bibliography{References.bib}

\end{document}